\documentclass[%
aip,
 jmp,%
 amsmath,amssymb,
preprint,%
]{revtex4-1}

\usepackage{graphicx}
\usepackage{bm}
\usepackage{amssymb, amsmath, amsbsy}
\usepackage{mathdots} 
\usepackage{mathrsfs} 
\usepackage{stackrel} 
\usepackage{graphicx}
\usepackage{dcolumn}

\usepackage{lipsum}
\usepackage{float}

\begin{document}

\title[Downconversion efficiency in MKIDs]{Interplay between phonon downconversion efficiency, density of states at Fermi energy, and intrinsic energy resolution for microwave kinectic inductance detectors. }

\author{Israel Hernandez}
\affiliation{Universidad de Guanajuato, Guanajuato, Mexico.}

\author{Gustavo Cancelo}
\author{Juan Estrada}
\author{Humberto Gonzalez}
\author{Andrew Lathrop}
\affiliation{Fermi National Accelerator Laboratory, Batavia, IL, USA.}

\author{Martin Makler}
\affiliation{Centro Brasileiro de Pesquisas Fisicas, Rio de Janeiro, Brazil.}

\author{Chris Stoughto}
\affiliation{Fermi National Accelerator Laboratory, Batavia, IL, USA.}

\date{\today}

\begin{abstract}
Microwave Kinetic Inductance detectors (MKIDs) have been recognized as a powerful new tool for single photon detection. These highly multiplexed superconducting devices give timing and energy measurement for every detected photon. However, the full potential of MKID single photon spectroscopy has not been reached , the achieved energy resolution is lower than expected from first principles. Here, we study the efficiency in the phonon downconversion process following the absorption of energetic photons in MKIDs. Assuming previously published material properties, we measure an average downconversion efficiency for three TiN resonators is $\eta$=0.14. We discuss how this efficiency can impact the intrinsic energy resolution of MKID, and how any uncertainty in the unknown density of electron states at the Fermi energy directly affects the efficiency estimations.
\end{abstract}

\maketitle

\section{MKIDs and photon energy resolution. }

Microwave kinetic inductance detectors (MKIDs) are multipixel superconducting photon detectors \cite{2003Nature}.  Each MKID pixel is a microwave resonator fabricated by depositing a superconducting thin film  over a substrate. The sensor structure is printed with standard microfabrication techniques. Photons hitting an MKID pixel deposit energy on the superconducting material breaking Cooper pairs and creating quasiparticles. The kinetic inductance of the superconductor depends on the density of Cooper pairs, and when a photon is absorbed in the MKID the kinetic inductance increases. The change in the inductance produces a shift in the resonance frequency with magnitude proportional to the deposited energy, and this change in resonance frequency is readout using a microwave feed line. The resonance frequency for each pixel can be adjusted by design, and arranged such that several (up to thousands) pixels can be frequency multiplexed into a single microwave feed line. Because of their mutiplexing potential, MKIDs  are being developed for millimeter wave astronomy \cite{Superscpec}, cosmic microwave background observations \cite{2018KIDSCMB}, optical and near-IR astronomy \cite{Gigaz} and dark matter detectors \cite{2008Golwala}.


The number of quasiparticles produced on MKIDs when a photon is detected can be expressed as
\begin{eqnarray}
    N_{qp}(\nu)= \frac{\eta h\nu}  {\Delta},
   \label{QuasiPhoton}
 \end{eqnarray}
 where $h$ is the Plank constant, $\nu$ is the frequency of the detected photon, and $\Delta$ is the superconducting energy gap. The low temperature limit for the critical temperature of the superconductor is  $T_c  = \Delta/ (1.75 k_B)$ from BCS\cite{Bardeen1957}, with  $k_B$ the Boltzman constant.  The downconversion efficiency $\eta$ is the fraction of the energy of the photon that is converted into quasiparticles and is calculated to be $\eta = 0.57$ based on Ref. \cite{downEff2000}, this value has been commonly used for the estimation of the intrinsic energy resolution for MKIDs. The efficiency $\eta$ quantifies any losses produced in the energy downconversion process involved in going from the $\sim$eV energy scale of the absorbed photon, to the $\Delta\sim 1$ meV energy of the quasiparticles.
 
The fundamental performance limit for the dispersion of the energy measurement in an MKID pixel is given by quasiparticle statistics
\begin{equation}
\sigma_{\nu} = \sqrt{F N_{qp}(\nu)},
\end{equation}
where $F \sim 0.2$ is the Fano factor \cite{Fano_facto} that accounts for the fact that the cascade process generating quasiparticles is highly correlated. The maximum energy resolution for photon detection is then
\begin{equation}
R_{\max}(\eta)=\frac{E_\gamma}{\Delta E_\gamma} =\frac{1}{2.355}\sqrt{\frac{h \eta \nu}{ F \Delta}}.
\label{eq:eres}
\end{equation}
For a blue photon with $E_\gamma= h \nu = 2.73$ eV 
($\lambda$~=~455~nm), in a TiN superconductor with $\Delta=1.3$ meV ($T_c~=~0.88$~K), and assuming $\eta=0.57$, the maximum energy resolution would be $R_{\max}=103$. 

The intrinsic resolution $R_{\max}(\eta=0.57)$ has not yet been achieved for any MKID sensor in the single photon counting regime. The summary of results achieved for visible and near-IR photon detection with MKIDs is presented in Table~\ref{tabl:resolutionLab}. The measured resolutions $R$ are typically an order of magnitude lower than $R_{\max}$, assuming $\eta=0.57$ in Eq.(\ref{eq:eres}).

\begin{table}
\centering
\begin{tabular}{||c | c | c | c| c | c||} 
\hline
$\lambda$ (nm) & $R$ & $R_{\max} (\eta=0.57)$  & Tc (K) & SC film & Ref \\ [0.5ex] 
 \hline\hline
 808  & 9.4 & 155 &  0.395  & Hf & \cite{MKIDHf2019} \\ 
 \hline
 808 & 8.9 &  75 & 0.930 & TiN & \cite{2019ParaAMP} \\
 
 \hline
 808 & 8.1 & 73  & 0.900 & PtSi & \cite{2017PtSi} \\
 \hline
 980 & 6.3 & 69  & 0.900 & PtSi & \cite{2017PtSi} \\
 \hline
1310 & 5.8 & 60  & 0.900 & PtSi & \cite{2017PtSi} \\
\hline
400 & 10 & 107  & 0.930 & TiN & \cite{2013Ben} \\
\hline
 808 & $23^{*}$ &  75 & 0.930 & TiN & \cite{2019ParaAMP} \\
 \hline
\end{tabular}
\caption{Single photon energy resolution measured in MKIDs developed for optical and near-IR astronomy, compared to the maximum energy resolution. The $R_{\max}$ values are calculated using Eq.(\ref{eq:eres}) for $\eta$=0.57 and $F=0.2$. $^*$The last line in this table is not a measurement, and corresponds to the expected energy resolution for an MKID when using a Parametric Amplifier.\cite{2019ParaAMP}}
\label{tabl:resolutionLab}
\end{table}

\begin{figure}
    \centering
    \includegraphics[width=0.8\columnwidth]{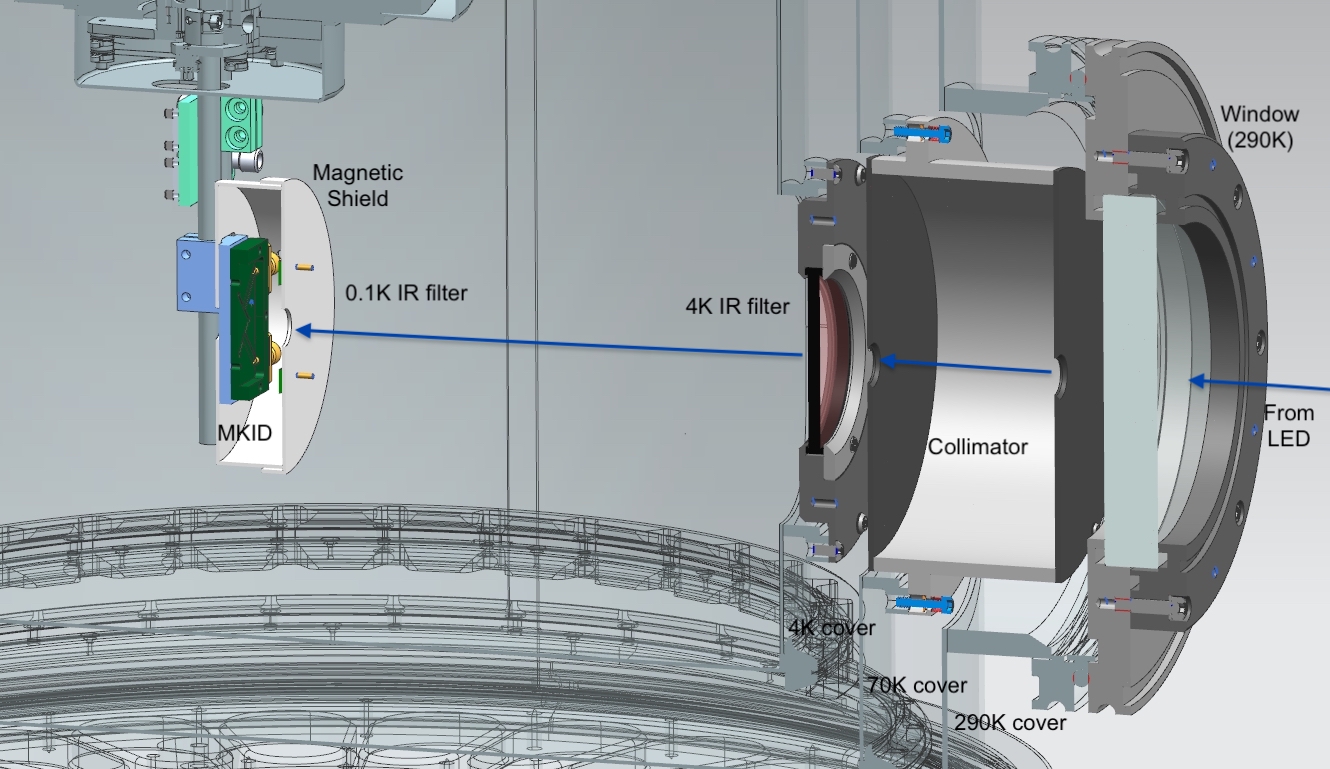}
    \includegraphics[width=0.5 \columnwidth
    ]{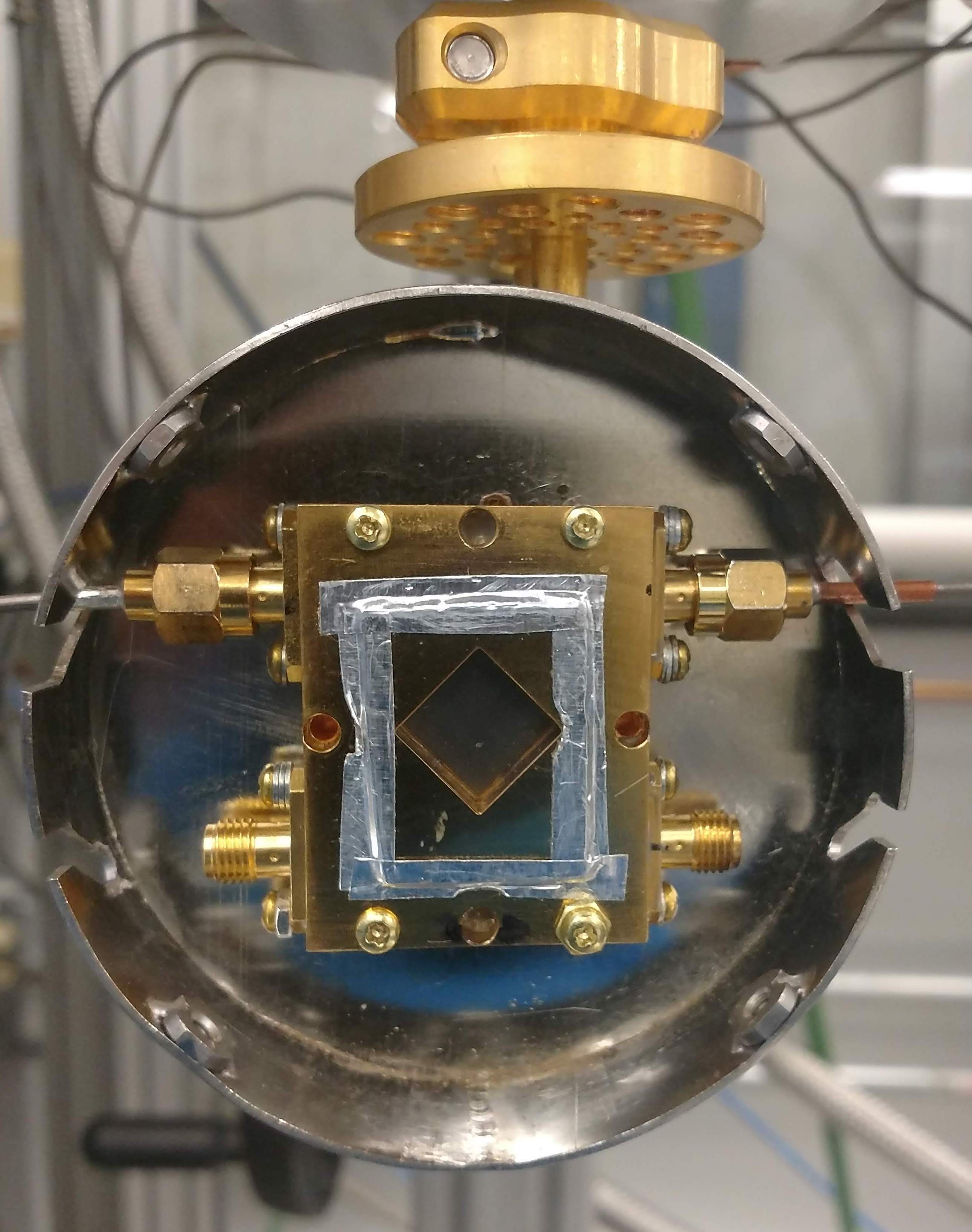}
    \caption{Top: Schematic of ADR refrigetator with MKID installed inside the cold stage using a magnetic shield. The window allows for broadband illumination with filtering to give a negligible thermal load from IR radiation. Bottom: MKID package with micromirror array developed by the UCSB group \cite{DARKNESS}.}
    \label{fig:ADRwindow}
\end{figure}

In this work we study the interplay between $\eta$ and  material properties, and its potential role in the degradation of the energy resolution for MKIDs. From Eq.(\ref{eq:eres}), the intrinsic energy resolution $R_{\max} (\eta) \propto {\eta}^{1/2}$. A technique is proposed to measure this efficiency, and applied to a few resonator pixels in a device fabricated with TiN. The MKID sensor used in this work is described in Ref.\cite{2013Ben}. The device has a total of  2024 ($44 \times 46$) pixels, with typical performance of  $R \sim 10$ at 400 nm. A microlens array is included in the MKID package to focus the incoming radiation into the inductor element of each pixel, as described in Ref.\cite{2013Ben} . The operations were performed in an Adiabatic Demagnetization Refrigerator (ADR) with a base temperature of 30 mK, and equipped with an optical window designed to allow illumination  while keeping a low thermal load (see Fig.\ref{fig:ADRwindow}).

\section{Quasiparticle density in thermal equilibrium}

The number of quasiparticles in thermal equilibrium as a function of temperature ($T$) is discussed in Ref.\cite{NqpTheoryBook} and can be expressed as
\begin{equation}
   N_{qp}(T)=2~VN_0\sqrt{2\pi k_{B} \Delta T}e^{-\frac{\Delta}{k_{B}T}}  ,
   \label{eq:quasithermal}
\end{equation}
where $V$ is the volume of the superconductor ($V~=~96~\mu$m$^3$ as designed, and could have fluctuations due to fabrication tolerances), and $N_0$ is the single-spin density of electron states at the Fermi energy of the metal. Gao et al.\cite{2012Gao} have estimated its value to be $N_0^{Gao}=3.9~10^{10}$eV$^{-1}~\mu$m$^{-3}$ making a strong assumption on phonon downconversion efficiency. We use this value for reference, express our results in terms of $\mathcal{X}=N_0/N_0^{Gao}$,
and later discuss  the importance of this parameter. 

\begin{figure}
        \centering \includegraphics[width=1.\columnwidth]{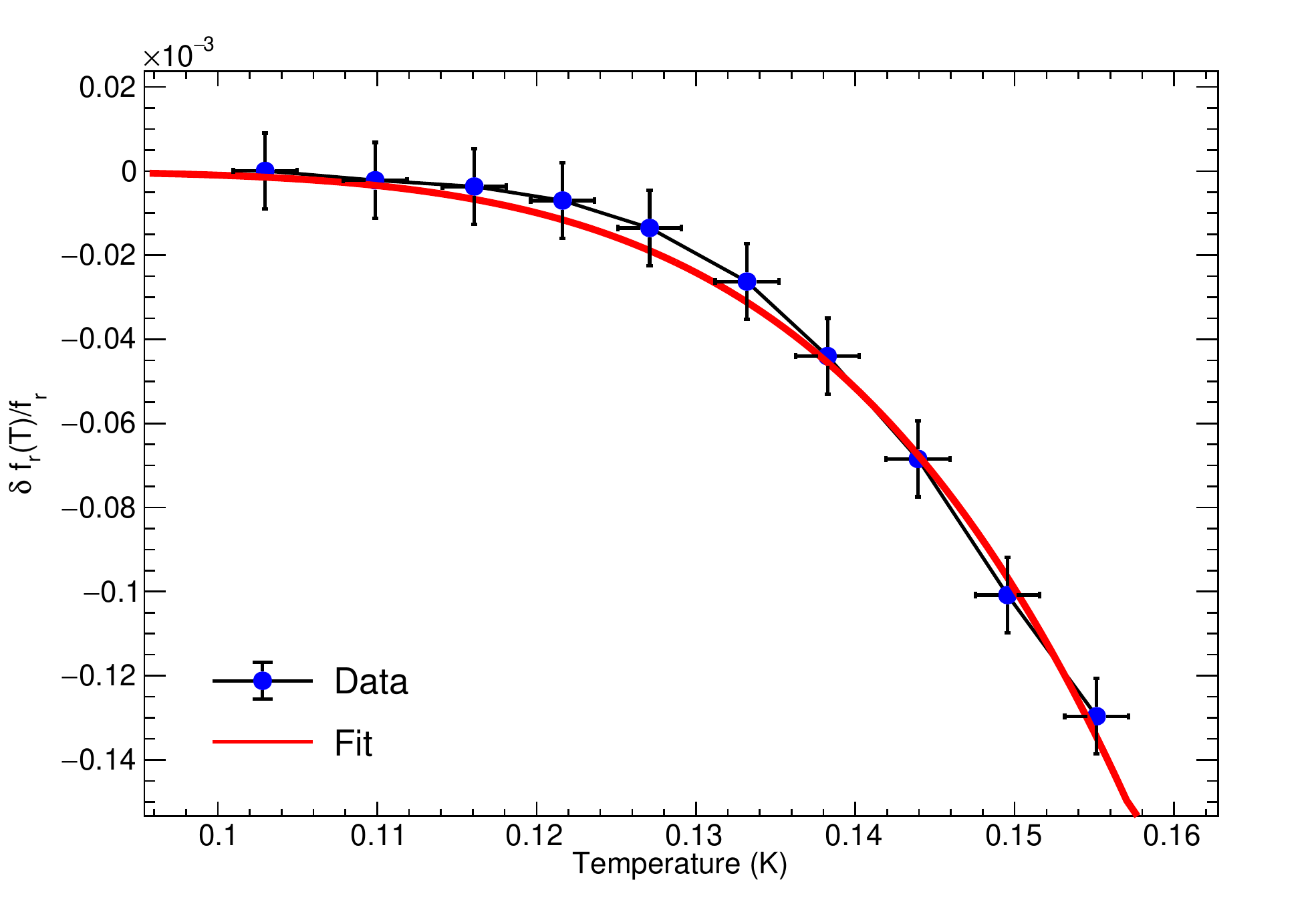}
        \caption{Fractional  frequency change $\delta f_{r}/f_{r}$ as a function of the temperature for one resonator. The red line is a fit to Eqs.(\ref{eq:frevstemp},\ref{eq:Mattis}). The errors in frequency are determined by comparing different techniques for fitting the resonance curve. The errors in temperature reflect a potential offset between our temperature sensor and the resonator.}
        \label{fig:frevstemp}
\end{figure}

By scanning the operating temperature of an MKID it is possible to change $N_{qp}$ in a controlled way. Since the kinetic inductance depends on $N_{qp}$, it is possible to calibrate the response of the MKID as a function of $N_{qp}$. Figure~\ref{fig:frevstemp} shows the change on the resonance frequency $f_{r}(T)$ for a TiN pixel as a function of temperature. Using the Mattis-Bardeen model \cite{Mattis1958} this can be written as
\begin{equation}
\frac{\delta f_{r}}{f_{r}}=\frac{f_{r}(T)-f_{r}(0)}{f_{r}(0)}=-\frac{1}{2}\alpha \frac{\delta \sigma_{2}(T)}{\sigma_{2}(0)},
\label{eq:frevstemp}
\end{equation}
where $f_{r}(0)$ is the low temperature limit  $T \sim 0$ K, $\alpha$ is constant,  and  $\sigma_{2}(T)$ is the imaginary part of the conductivity  given by
\begin{equation} 
\frac{\sigma_{2}(T)}{\sigma_{N}} \simeq \frac{\pi\Delta}{\hbar\omega}\left[1-2\exp\left(-\frac{\Delta}{k_{b}T}\right)\exp\left(-\frac{\hbar\omega}{2k_{b}T}\right)I_{0}\left(\frac{\hbar\omega}{2k_{b}T}\right)\right]  .
\label{eq:Mattis}
\end{equation}
Here, $\sigma_{N}$ the normal state resistance. The superconducting energy gap $\Delta$ (and therefore the critical temperature) can be determined by fitting Eq.~\ref{eq:frevstemp} to the data in Fig.~\ref{fig:frevstemp}. The result for this resonator is $T_c = 0.88 \pm 0.04$ K.

\begin{figure}
\includegraphics[width=1.\columnwidth]{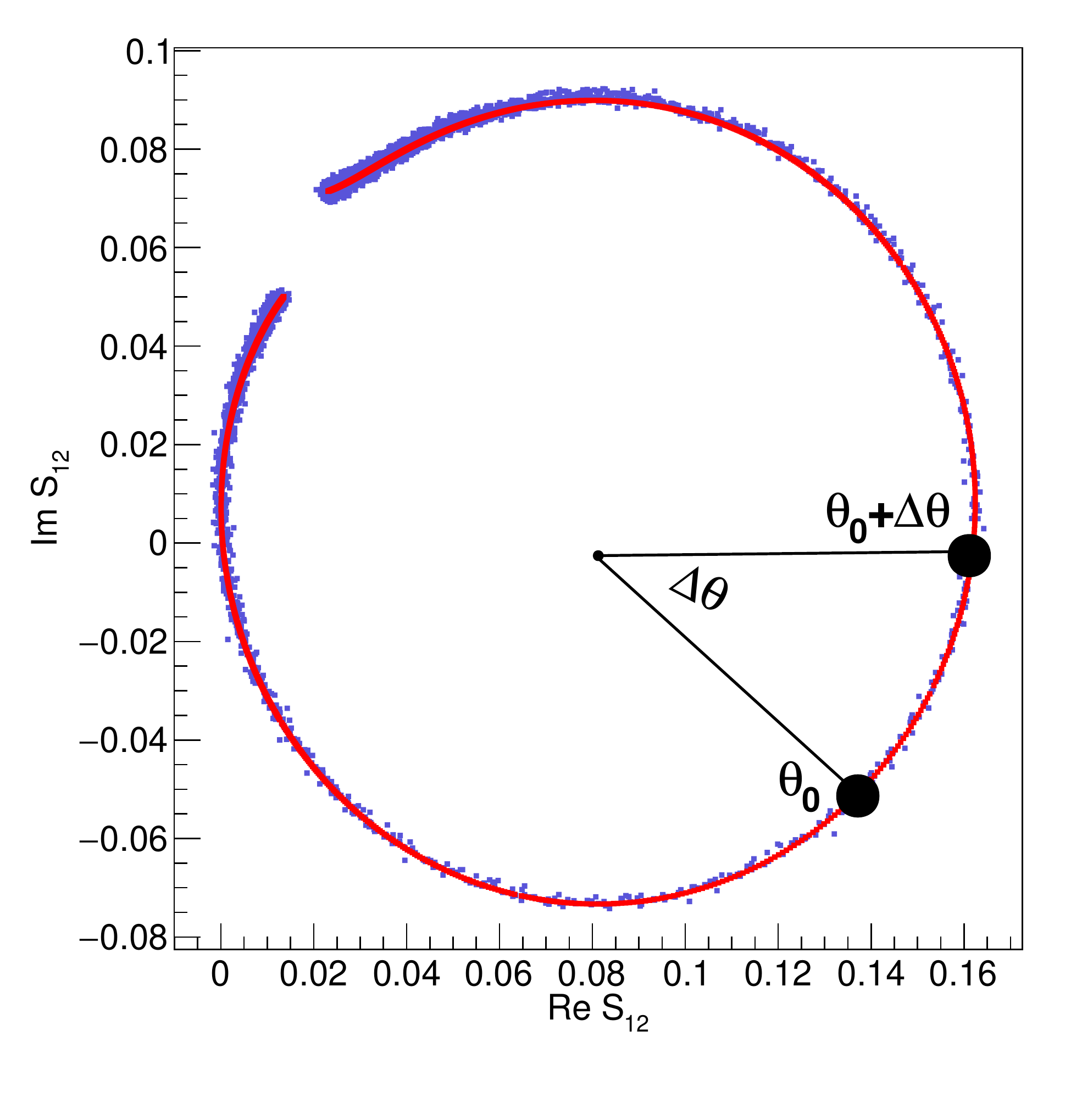}
\caption{Complex transmission amplitude measured for a single MKID pixel $S_{21}(f)$. For the resonance frequency $S_{21}(f_r)=A~e^{i\theta_0}$ . When the number of quasiparticles changes due to a change in temperature, or the absorption of a photon, the phase rotates as $S_{21}(f_r)=A~e^{i\theta_0+\Delta \theta}$.}
\label{fig:ComplexS12}
\end{figure}

\begin{figure}
    \centering
    \includegraphics[width=1.0\columnwidth]{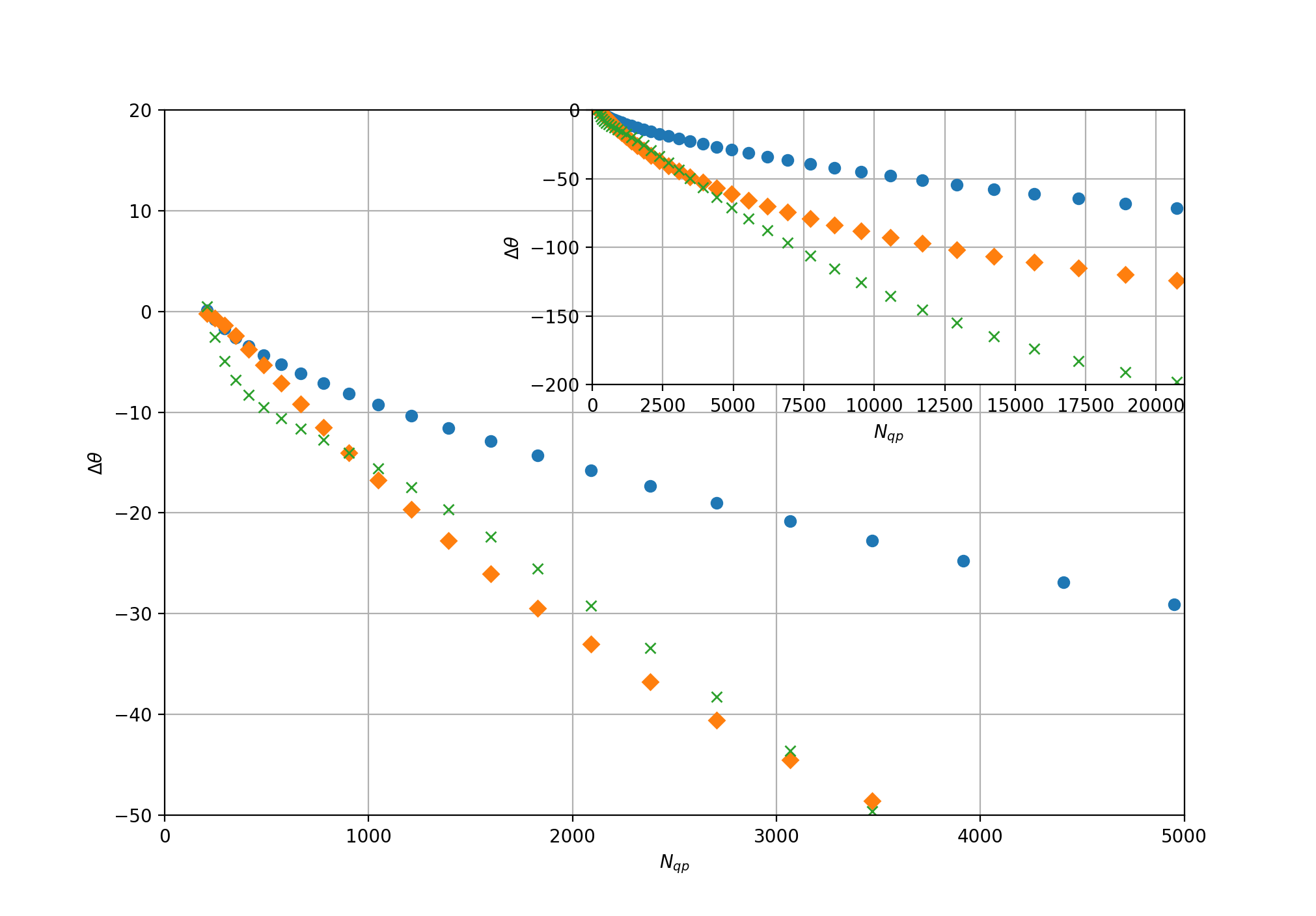}
    \caption{Phase rotation as a function of quasiparticle number (assuming $/\mathcal{X}=1$) for three pixels in the TiN MKID. Blue, orange and green corresponds to pixel (a),(b) and (c). The inset shows the same data in an extended range. }
    \label{fig:thetaNqp}
\end{figure}

For practical reasons the readout of the MKID photosensor is not done looking at the change in resonance frequency directly,  it is rather done measuring the phase $\theta$ of the signal transmission  $S_{21}(f_r)$ in the complex plane at a fixed frequency using the tools discussed in Ref.\cite{MKIDreadout2012}, Ref.\cite{DARKNESS2018} and Ref.\cite{Isra2019}. The frequency $f_r$ is selected as the resonant frequency when the pixel is dark and in thermal equilibrium. Figure~\ref{fig:ComplexS12} shows the transmission in the complex plane, where the larger black dots mark  $f_r$. When the temperature changes, the phase angle for $S_{21}(f_r)$ rotates, as indicated in Fig.~\ref{fig:ComplexS12}. For each temperature, we measure the change in the phase at the resonance frequency $\Delta \theta$. Through  Eq.(\ref{eq:quasithermal}) this rotation can be expressed as a function of $\mathcal{N}_{qp} = N_{qp}/\mathcal{X}$ as shown in Fig.~\ref{fig:thetaNqp} for three pixels in the array labeled (a),(b) and (c). Therefore a rotation in phase $\Delta \theta$ can be directly associated with a change in the number of quasiparticles  $N_{qp}$.

\section{MKIDs response to photons.}

\begin{figure}
    \centering
    \includegraphics[width=1.0\columnwidth]{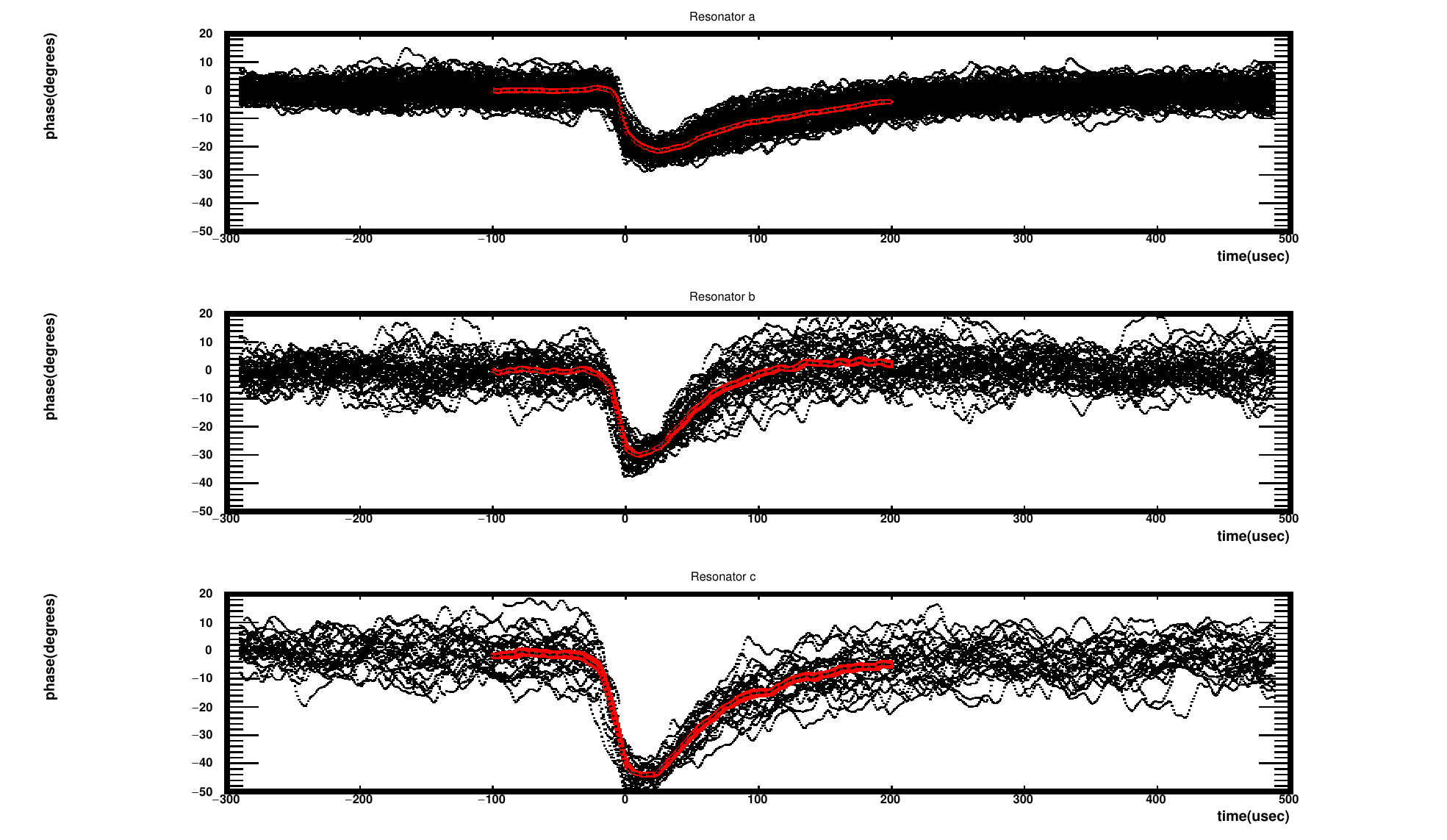}
    \caption{Phase rotation pulses produced by the detection of a blue photon ($E_\gamma = 2.73$ eV) in three different MKID pixels. Each black curve corresponds to a single pulse,  athe verage $\sim100$ pulses is shown in red. Pixels a, b and c from top to bottom. The shape of the pulses shown here could be affected by non-optimal signal processing, and this systematic effects will be studied in future work.}
    \label{fig:photonPulse}
\end{figure}

In the previous section the MKID response is calibrated for a controlled change in $\mathcal{N}_{pq}$ by varying the temperature.  In this section we measure the responsivity of the MKID pixel to blue photons ($E_{\gamma}=2.73$ eV). The rotation in the phase produced by the photon is measured using the  readout system described in Ref.\cite{Arcons2024} and the results are shown in Fig.~\ref{fig:photonPulse} for the pixels (a), (b) and (c). The photons produce a phase pulse, with a sharp rise time and about $100~\mu$s decay time.  The curves in Fig.\ref{fig:photonPulse} shoe the phase shift as a function of time, averaged over $\sim$100 monochromatic photon pulses detected on each pixel. The phase rotation $\Delta \theta$ is measured at the time corresponding to minimum of the curve, around time $\sim 25~\mu$s after the pulse.  The values for each pixel are shown in the second column of Table \ref{tabl:deltaphi}. The observed response to monoenergetic photons can be combined with the curve in Fig.\ref{fig:thetaNqp} to estimate the observed change in the number of quasiparticles and the results are presented in Table \ref{tabl:deltaphi}. The observed downconversion efficiency is given by Eq.(\ref{QuasiPhoton}), such that
\begin{equation}
    \frac{\eta(E_{\gamma})}{\mathcal{X}}=  \frac{ \mathcal{N}_{pq}(E_{\gamma}) }{E_{\gamma}/\Delta} ,
\end{equation}
and the results are shown in the fourth columns of         Table \ref{tabl:deltaphi}.
The observed phonon downconversion efficiency is about a factor of 4 lower than the expected for bulk material \cite{downEff2000}, giving a significant reduction of the intrinsic resolution compared to that expected for bulk materials.

\begin{table}
\centering
\begin{tabular}{||c | c | c | c | c ||} 
 \hline
 pixel & $\Delta\theta$ (deg.) & $\mathcal{N}_{qp}/\mathcal{X}$ & $\eta_{}/\mathcal{X}$   & $R_{\max}/\sqrt{\mathcal{X}}$\\ [0.5ex] 
 \hline\hline
a  & 22 & 3314 &  0.17   & 56 \\ 
b  & 30 & 1869 &  0.09   & 41 \\ 
c  & 44 & 3093 &  0.15   & 52 \\ 
 \hline
 \hline
 \end{tabular} 
 \caption{Single photon energy resolution measured MKIDs developed for optical and near-IR astronomy, compared with the the maximum energy resolution. The $R_{\max}/\sqrt{\mathcal{X}}$ values are calculated using Eq.(\ref{eq:eres}) for $\eta$ and $F=0.2$.
\cite{2019ParaAMP}}
\label{tabl:deltaphi}
\end{table}

\section{Conclusion}

The phase rotation in MKIDs as a function of the number of quasiparticles was measured by studying devices in thermal equilibrium. The phase rotation pulse was also measured for MKIDs detecting single visible photons ($\lambda = 455$ nm). By comparing the two observations, the number of quasiparticles produced by the photon detection is determined, providing a direct measurement of the photon downconversion efficiency. The average efficiency for the three pixels used in this study is $\eta / \mathcal{X}=0.14$. Assuming  $\mathcal{X}\sim1$ following the estimation by Gao et al.\cite{2012Gao}, this points to a downconversion efficiency $\eta$ a factor of 4 lower than the value expected for bulk material. This reduced value of $\eta$ will produce a degradation in the photon energy resolution for the device by a factor of $\sim$ 2, according to  Eq.(\ref{eq:eres}). 

In order to understand the intrinsic energy resolution of MKIDs, it is essential to determine the down conversion efficiency $\eta$ and the density of states at the Fermi energy $N_0$ (parameterized by $\mathcal{X}$ in this work).  The value of $N_0^{Gao}$  used as reference here is based on an estimation from a photon sensitivity measurement assuming $\eta=0.6$ for devices with somewhat higher critical temperature\cite{2012Gao}.
$N_0$ is estimated from first principles for similar films as $N_0=8.9~10^{9}$eV$^{-1}~\mu$m$^{-3}$ in Leduc et al.\cite{Leduc2010} based on calculation by Dridi et al.\cite{Dridi2002}, these numbers are significantly lower than our reference and would indicate $\mathcal{X}\sim0.2$, resulting in a lower downconversion efficiency. 

It has recently been demonstrated that the energy resolution for MKIDs can be degraded by noise from the cryogenic low temperature amplifier in the readout system \cite{2019ParaAMP}. Using a wide-band parametric amplifier it is expected that the energy resolution will increase significantly to $R/R_{\mbox{max}}=3$ as shown in Table \ref{tabl:resolutionLab} according to Ref.\cite{2019ParaAMP}. Considering the reduction for $R_{\mbox{max}}$ in Eq.(\ref{eq:eres}) for $\eta=0.15$, the expected energy resolution for readout with parametric amplifier would be within $\sim 60$\% of the intrinsic energy resolution.

A model for photon downconversion efficiency for thin films is discussed Ref.\cite{quasiPartDown2014}. The model indicates that the efficiency for producing quasiparticles depends strongly on the ratio of the phonon loss characteristic time $\tau_l$, to the photon life time for the superconductor $\tau_0$. The model seems to indicate that for the measured $\eta=0.14$, this ratio is $\tau_l/\tau_0< 0.5$. A significant improvement in the performance of the MKID would be achieved by reducing the phonon losses in the thin film.

Phonon losses during downconversion could be controlled by designing intefaces which keep the phonons inside the superconducting film as discussed in Ref.\cite{Kaplan1979}, or by otherwise optimizing the geometry of the devices. We leave for future effort the exploration of these ideas to increase $\eta$, making it closer to the values expected for bulk material.

\section*{Acknowledgements}

 We thank Prof. Ben Mazin for his support providing the sensors used for this tests, and giving advice about the operation of these devices. We also thank Prof. Mazin for important comments pointing to the future work needed to fully understand the issues discussed here. We thank the team at the Fermilab Silicon Detector facility for their support in the operations of the Adiabatic Demagnetization Refrigerator used in this experiment. We specially thank Donna Kubik and Jorge Montes for their support. This work is completed as part of the requirements for the Master thesis of Israel Hernandez at the University of Guanajuato. MM is partially supported by CNPq and FAPERJ. Fora Bozo


\subsection{Citations and References} \label{sec:citeref}

\bibliographystyle{JHEP}
\bibliography{MKIDbibliop}

\providecommand{\href}[2]{#2}\begingroup\raggedright\begin{thebibliography}{10}

\bibitem{2003Nature}
P.~K. {Day}, H.~G. {LeDuc}, B.~A. {Mazin}, A.~{Vayonakis}, and J.~{Zmuidzinas},
  {\it {A broadband superconducting detector suitable for use in large
  arrays}},  {\em nat} {\bf 425} (Oct, 2003) 817--821.

\bibitem{Superscpec}
E.~{Shirokoff}, P.~S. {Barry}, C.~M. {Bradford}, G.~{Chattopadhyay}, P.~{Day},
  S.~{Doyle}, S.~{Hailey-Dunsheath}, M.~I. {Hollister}, A.~{Kov{\'a}cs},
  C.~{McKenney}, H.~G. {Leduc}, N.~{Llombart}, D.~P. {Marrone}, P.~{Mauskopf},
  R.~{O'Brient}, S.~{Padin}, T.~{Reck}, L.~J. {Swenson}, and J.~{Zmuidzinas},
  {\it {MKID development for SuperSpec: an on-chip, mm-wave, filter-bank
  spectrometer}},  in {\em Millimeter, Submillimeter, and Far-Infrared
  Detectors and Instrumentation for Astronomy VI}, vol.~8452 of {\em procspie},
  p.~84520R, Sept., 2012.
\newblock \href{http://arxiv.org/abs/1211.1652}{{\tt arXiv:1211.1652}}.

\bibitem{2018KIDSCMB}
P.~S. {Barry}, S.~{Doyle}, A.~L. {Hornsby}, A.~{Kofman}, E.~{Mayer},
  A.~{Nadolski}, Q.~Y. {Tang}, J.~{Vieira}, and E.~{Shirokoff}, {\it {Design
  and Performance of the Antenna-Coupled Lumped-Element Kinetic Inductance
  Detector}},  {\em Journal of Low Temperature Physics} {\bf 193} (Nov, 2018)
  176--183, [\href{http://arxiv.org/abs/1801.06265}{{\tt arXiv:1801.06265}}].

\bibitem{Gigaz}
D.~W. {Marsden}, B.~A. {Mazin}, K.~{O'Brien}, and C.~{Hirata}, {\it {Giga-z: A
  100,000 Object Superconducting Spectrophotometer for LSST Follow-up}},  {\em
  apjs} {\bf 208} (Sep, 2013) 8, [\href{http://arxiv.org/abs/1307.5066}{{\tt
  arXiv:1307.5066}}].

\bibitem{2008Golwala}
S.~{Golwala}, J.~{Gao}, D.~{Moore}, B.~{Mazin}, M.~{Eckart}, B.~{Bumble},
  P.~{Day}, H.~G. {Leduc}, and J.~{Zmuidzinas}, {\it {A WIMP Dark Matter
  Detector Using MKIDs}},  {\em Journal of Low Temperature Physics} {\bf 151}
  (Apr, 2008) 550--556.

\bibitem{Bardeen1957}
J.~Bardeen, L.~N. Cooper, and J.~R. Schrieffer, {\it Theory of
  superconductivity},  {\em Phys. Rev.} {\bf 108} (Dec, 1957) 1175--1204.

\bibitem{downEff2000}
A.~G. Kozorezov, A.~F. Volkov, J.~K. Wigmore, A.~Peacock, A.~Poelaert, and
  R.~den Hartog, {\it Quasiparticle-phonon downconversion in nonequilibrium
  superconductors},  {\em Phys. Rev. B} {\bf 61} (May, 2000) 11807--11819.

\bibitem{Fano_facto}
U.~Fano, {\it Ionization yield of radiations. ii. the fluctuations of the
  number of ions},  {\em Phys. Rev.} {\bf 72} (Jul, 1947) 26--29.

\bibitem{MKIDHf2019}
N.~{Zobrist}, G.~{Coiffard}, B.~{Bumble}, N.~{Swimmer}, S.~{Steiger},
  M.~{Daal}, G.~{Collura}, A.~B. {Walter}, C.~{Bockstiegel}, N.~{Fruitwala},
  I.~{Lipartito}, and B.~A. {Mazin}, {\it {Design and performance of hafnium
  optical and near-IR kinetic inductance detectors}},  {\em Applied Physics
  Letters} {\bf 115} (Nov, 2019) 213503,
  [\href{http://arxiv.org/abs/1911.06434}{{\tt arXiv:1911.06434}}].

\bibitem{2019ParaAMP}
N.~{Zobrist}, B.~H. {Eom}, P.~{Day}, B.~A. {Mazin}, S.~R. {Meeker},
  B.~{Bumble}, H.~G. {LeDuc}, G.~{Coiffard}, P.~{Szypryt}, N.~{Fruitwala},
  I.~{Lipartito}, and C.~{Bockstiegel}, {\it {Wide-band parametric amplifier
  readout and resolution of optical microwave kinetic inductance detectors}},
  {\em Applied Physics Letters} {\bf 115} (Jul, 2019) 042601,
  [\href{http://arxiv.org/abs/1907.03078}{{\tt arXiv:1907.03078}}].

\bibitem{2017PtSi}
P.~{Szypryt}, S.~R. {Meeker}, G.~{Coiffard}, N.~{Fruitwala}, B.~{Bumble},
  G.~{Ulbricht}, A.~B. {Walter}, M.~{Daal}, C.~{Bockstiegel}, G.~{Collura},
  N.~{Zobrist}, I.~{Lipartito}, and B.~A. {Mazin}, {\it {Large-format platinum
  silicide microwave kinetic inductance detectors for optical to near-IR
  astronomy}},  {\em Optics Express} {\bf 25} (Oct, 2017) 25894,
  [\href{http://arxiv.org/abs/1710.07318}{{\tt arXiv:1710.07318}}].

\bibitem{2013Ben}
B.~A. {Mazin}, S.~R. {Meeker}, M.~J. {Strader}, P.~{Szypryt}, D.~{Marsden},
  J.~C. {van Eyken}, G.~E. {Duggan}, A.~B. {Walter}, G.~{Ulbricht},
  M.~{Johnson}, B.~{Bumble}, K.~{O'Brien}, and C.~{Stoughton}, {\it {ARCONS: A
  2024 Pixel Optical through Near-IR Cryogenic Imaging Spectrophotometer}},
  {\em pasp} {\bf 125} (Nov, 2013) 1348,
  [\href{http://arxiv.org/abs/1306.4674}{{\tt arXiv:1306.4674}}].

\bibitem{DARKNESS}
S.~R. {Meeker}, B.~A. {Mazin}, A.~B. {Walter}, P.~{Strader}, N.~{Fruitwala},
  C.~{Bockstiegel}, P.~{Szypryt}, G.~{Ulbricht}, G.~{Coiffard}, B.~{Bumble},
  G.~{Cancelo}, T.~{Zmuda}, K.~{Treptow}, N.~{Wilcer}, G.~{Collura},
  R.~{Dodkins}, I.~{Lipartito}, N.~{Zobrist}, M.~{Bottom}, J.~C. {Shelton},
  D.~{Mawet}, J.~C. {van Eyken}, G.~{Vasisht}, and E.~{Serabyn}, {\it
  {DARKNESS: A Microwave Kinetic Inductance Detector Integral Field
  Spectrograph for High-contrast Astronomy}},  {\em PASP} {\bf 130} (June,
  2018) 065001, [\href{http://arxiv.org/abs/1803.10420}{{\tt
  arXiv:1803.10420}}].

\bibitem{NqpTheoryBook}
G.~Lehmann, {\it Nonequilibrium superconductivity, phonons and kapitza
  boundaries},  {\em Crystal Research and Technology} {\bf 17} (1982), no.~8
  1047--1047,
  [\href{http://arxiv.org/abs/https://onlinelibrary.wiley.com/doi/pdf/10.1002/crat.2170170828}{{\tt
  https://onlinelibrary.wiley.com/doi/pdf/10.1002/crat.2170170828}}].

\bibitem{2012Gao}
J.~{Gao}, M.~R. {Vissers}, M.~O. {Sandberg}, F.~C.~S. {da Silva}, S.~W. {Nam},
  D.~P. {Pappas}, D.~S. {Wisbey}, E.~C. {Langman}, S.~R. {Meeker}, B.~A.
  {Mazin}, H.~G. {Leduc}, J.~{Zmuidzinas}, and K.~D. {Irwin}, {\it {A
  titanium-nitride near-infrared kinetic inductance photon-counting detector
  and its anomalous electrodynamics}},  {\em Applied Physics Letters} {\bf 101}
  (Oct, 2012) 142602, [\href{http://arxiv.org/abs/1208.0871}{{\tt
  arXiv:1208.0871}}].

\bibitem{Mattis1958}
D.~C. Mattis and J.~Bardeen, {\it Theory of the anomalous skin effect in normal
  and superconducting metals},  {\em Phys. Rev.} {\bf 111} (Jul, 1958)
  412--417.

\bibitem{MKIDreadout2012}
S.~{McHugh}, B.~A. {Mazin}, B.~{Serfass}, S.~{Meeker}, K.~{O'Brien}, R.~{Duan},
  R.~{Raffanti}, and D.~{Werthimer}, {\it {A readout for large arrays of
  microwave kinetic inductance detectors}},  {\em Review of Scientific
  Instruments} {\bf 83} (Apr, 2012) 044702--044702,
  [\href{http://arxiv.org/abs/1203.5861}{{\tt arXiv:1203.5861}}].

\bibitem{DARKNESS2018}
S.~R. {Meeker}, B.~A. {Mazin}, A.~B. {Walter}, P.~{Strader}, N.~{Fruitwala},
  C.~{Bockstiegel}, P.~{Szypryt}, G.~{Ulbricht}, G.~{Coiffard}, B.~{Bumble},
  G.~{Cancelo}, T.~{Zmuda}, K.~{Treptow}, N.~{Wilcer}, G.~{Collura},
  R.~{Dodkins}, I.~{Lipartito}, N.~{Zobrist}, M.~{Bottom}, J.~C. {Shelton},
  D.~{Mawet}, J.~C. {van Eyken}, G.~{Vasisht}, and E.~{Serabyn}, {\it
  {DARKNESS: A Microwave Kinetic Inductance Detector Integral Field
  Spectrograph for High-contrast Astronomy}},  {\em pasp} {\bf 130} (Jun, 2018)
  065001, [\href{http://arxiv.org/abs/1803.10420}{{\tt arXiv:1803.10420}}].

\bibitem{Isra2019}
I.~{Hernandez}, M.~{Makler}, J.~{Estrada}, C.~R. {Bom}, D.~{Kubik},
  J.~{Amette}, J.~{Montes}, and A.~{Leptop}, {\it {On the Homogeneity of TiN
  Kinetic Inductance Detectors Produced through Atomic Layer Deposition}},
  {\em arXiv e-prints} (Nov, 2019) arXiv:1911.06419,
  [\href{http://arxiv.org/abs/1911.06419}{{\tt arXiv:1911.06419}}].

\bibitem{Arcons2024}
B.~A. {Mazin}, S.~R. {Meeker}, M.~J. {Strader}, P.~{Szypryt}, D.~{Marsden},
  J.~C. {van Eyken}, G.~E. {Duggan}, A.~B. {Walter}, G.~{Ulbricht},
  M.~{Johnson}, B.~{Bumble}, K.~{O'Brien}, and C.~{Stoughton}, {\it {ARCONS: A
  2024 Pixel Optical through Near-IR Cryogenic Imaging Spectrophotometer}},
  {\em PASP} {\bf 125} (Nov., 2013) 1348,
  [\href{http://arxiv.org/abs/1306.4674}{{\tt arXiv:1306.4674}}].

\bibitem{Leduc2010}
H.~G. Leduc, B.~Bumble, P.~K. Day, B.~H. Eom, J.~Gao, S.~Golwala, B.~A. Mazin,
  S.~McHugh, A.~Merrill, D.~C. Moore, O.~Noroozian, A.~D. Turner, and
  J.~Zmuidzinas, {\it Titanium nitride films for ultrasensitive microresonator
  detectors},  {\em Applied Physics Letters} {\bf 97} (2010), no.~10 102509,
  [\href{http://arxiv.org/abs/https://doi.org/10.1063/1.3480420}{{\tt
  https://doi.org/10.1063/1.3480420}}].

\bibitem{Dridi2002}
Z.~Dridi, B.~Bouhafs, P.~Ruterana, and H.~Aourag, {\it First-principles
  calculations of vacancy effects on structural and electronic properties of
  {TiCxand} {TiNx}},  {\em Journal of Physics: Condensed Matter} {\bf 14} (oct,
  2002) 10237--10249.

\bibitem{quasiPartDown2014}
T.~{Guruswamy}, D.~J. {Goldie}, and S.~{Withington}, {\it {Quasiparticle
  generation efficiency in superconducting thin films}},  {\em Superconductor
  Science Technology} {\bf 27} (May, 2014) 055012,
  [\href{http://arxiv.org/abs/1401.1937}{{\tt arXiv:1401.1937}}].

\bibitem{Kaplan1979}
S.~B. {Kaplan}, {\it {Acoustic matching of superconducting films to
  substrates}},  {\em Journal of Low Temperature Physics} {\bf 37} (Nov., 1979)
  343--365.

\end{thebibliography}\endgroup
\setcitestyle{square}
\end{document}